\documentclass[conference]{IEEEtran}
\IEEEoverridecommandlockouts
\usepackage{cite}
\usepackage{amsmath,amssymb,amsfonts}
\usepackage{algorithm}
\usepackage{array}
\usepackage{mathrsfs}

\usepackage{textcomp}
\usepackage{stfloats}
\usepackage{url}
\usepackage{verbatim}
\usepackage{graphicx}
\usepackage{cite}
\usepackage{algpseudocode}
\usepackage{mathrsfs}
\usepackage{xcolor,soul}
\usepackage{adjustbox}
\usepackage{caption}
\usepackage{subcaption}

\newcommand{\ignore}[1]{}

\def\BibTeX{{\rm B\kern-.05em{\sc i\kern-.025em b}\kern-.08em
    T\kern-.1667em\lower.7ex\hbox{E}\kern-.125emX}}
\begin{document}
\title{Sparsity-Aware Optimization of In-Memory Bayesian Binary Neural Network Accelerators}
\author{\IEEEauthorblockN{Prabodh Katti, Bashir M. Al-Hashimi, Bipin Rajendran}
\IEEEauthorblockA{{Department of Engineering},
King's College London, London, UK \\
Email: bipin.rajendran@kcl.ac.uk}
}

\maketitle

\begin{abstract}
Bayesian Neural Networks (BNNs) provide principled estimates of model and data uncertainty by encoding parameters as distributions. This makes them key enablers for reliable AI that can be deployed on safety critical edge systems. These systems can be made resource efficient by restricting synapses to two synaptic states $\{-1,+1\}$ and using a memristive in-memory computing (IMC) paradigm. However, BNNs pose an additional challenge -- they require multiple instantiations for ensembling, consuming extra resources in terms of energy and area. In this work, we propose a novel sparsity-aware optimization for Bayesian Binary Neural Network (BBNN) accelerators that exploits the inherent BBNN sampling sparsity -- most of the network is made up of synapses that have a high probability of being fixed at $\pm1$ and require no sampling. The optimization scheme proposed here exploits the sampling sparsity that exists both among layers, i.e only a few layers of the network contain a majority of the probabilistic synapses, as well as the parameters i.e., a tiny fraction of parameters in these layers require sampling, reducing total sampled parameter count further by up to $86\%$. We demonstrate no loss in accuracy or uncertainty quantification performance for a VGGBinaryConnect network on CIFAR-100 dataset mapped on a custom sparsity-aware phase change memory (PCM) based IMC simulator. We also develop a simple drift compensation technique to demonstrate robustness to drift-induced degradation. Finally, we project latency, energy, and area for sparsity-aware BNN implementation in both pipelined and non-pipelined modes. With sparsity-aware implementation, we estimate upto $5.3 \times$ reduction in area and $8.8\times$ reduction in energy compared to a non-sparsity-aware implementation. Our approach also results in $2.9 \times $ more power efficiency compared to the state-of-the-art BNN accelerator. 

\end{abstract}


\section{Introduction}
 Bayesian neural networks (BNNs) offer a key advantage by quantifying uncertainty in their predictions, which makes them more reliable, especially for risk-sensitive applications like diagnostics, surveillance, and autonomous vehicles \cite{shukla2020mc,blundell2015weight}. This prevents overconfident decisions, which is crucial in high-risk environments \cite{Kendall_Gal_2017}. BNNs encode epistemic uncertainty by treating each parameter as a probability distribution, requiring ensemble predictors \cite{wilson2020case,jang2021bisnn}. However, this inference process is hardware-intensive as multiple instantiations are required for sampling \cite{katti2024bayesian, lu2022algorithm}. This paper explores efficiency gains by leveraging the inherent sparsity in Bayesian Binary Neural Networks, reducing the resource demands of ensembling.

 \noindent \emph{Related Work:} BNN accelerators on CMOS and FPGA have been explored \cite{10.1109/DAC18074.2021.9586137, cai2018vibnn}, but they face the `von Neumann bottleneck', leading to high energy, area consumption, and latency due to excessive data movement \cite{wulf1995hitting}. In-memory computing addresses this by co-locating memory and processing, using crossbars of non-volatile devices like PCM, RRAM, STT-RAM, and SOT-MRAM \cite{yang2020all, lu2022algorithm}, enabling efficient matrix-vector multiplication through Ohm's and Kirchhoff's laws.
\begin{figure}
\centering
\begin{subfigure}{0.18\textwidth}
 \centering   
 \includegraphics[width=\textwidth]{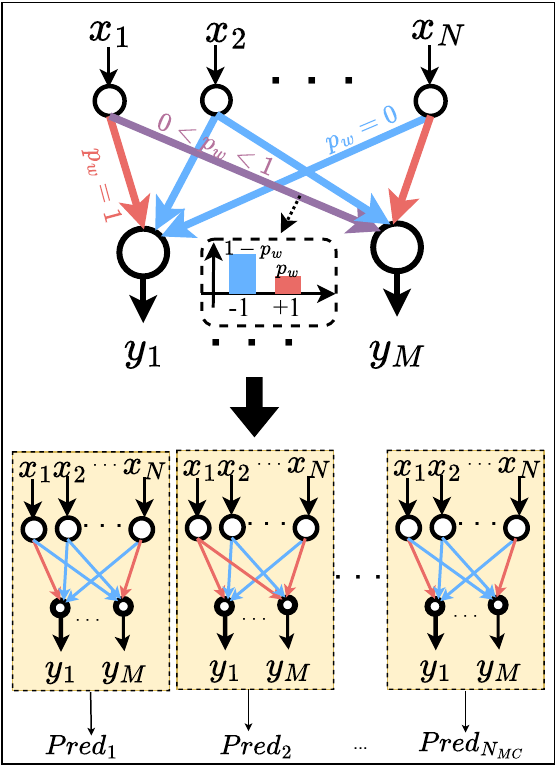}

\end{subfigure}
\hfill
\begin{subfigure}{0.14\textwidth}
   \centering \includegraphics[width=\textwidth]{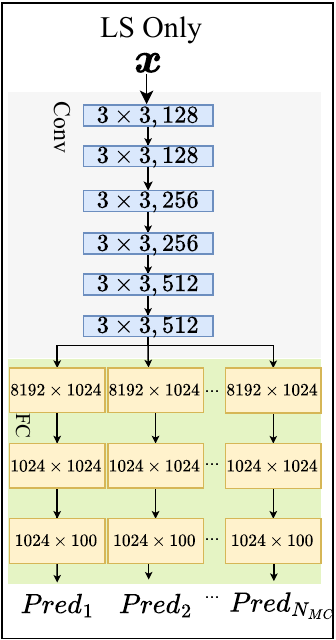}
\end{subfigure}
\hfill
\begin{subfigure}{0.10\textwidth}
   \centering \includegraphics[width=\textwidth]{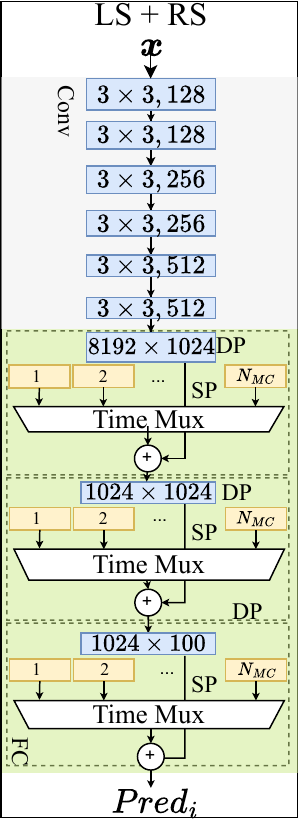}

\end{subfigure}
\caption {\textbf{Left}: Illustration of BBNN inference, where only a small fraction of synapses are probabilistic and thus participate in sampling. \textbf{Center}: Sparsity-aware optimization to exploit the layer sparsity of VGGBinaryConnect (only synaptic layers shown here). Only layers with a significant concentration of probabilistic synapses are utilized to create an ensemble of predictions, all available simultaneously. \textbf{Right}: Utilizing parameter sampling sparsity within these layers by separating rows with probabilistic synapses into `Stochastic Plane (SP)', and the rest into `Deterministic Plane (DP)'. The DP and one of the SP ensembles provide one set of predictions. Thus, predictions are available one by one.} 
\label{fig:sparsity-intro}
\end{figure}

RRAM, PCM, STT-RAM, and SOT-MRAM have been used  to implement Bayesian accelerators with Gaussian parameters 
\cite{bonnet2023bringing, 10181438,yang2020all,lu2022algorithm}.  Nanoscale device stochasticity has also been considered as a resource to generate noise for sampling using the central limit theorem, though requiring additional hardware \cite{yang2020all,lu2022algorithm}. In addition, these methods require Gaussian mean and standard deviation values to be stored per parameter. Architecture combining two devices to generate a Gaussian weight, co-locating entropy source and parameter storage was proposed in \cite{bonnet2023bringing}. However, the multi-level conductance involved introduces programming noise and requires reprogramming for drift mitigation. Binary parameter storage with Monte-Carlo dropout-based inference\cite{pmlr-v48-gal16} was used in \cite{ahmed2023spindrop}, which while easier to train, are less expressive for out-of-distribution (OOD) data \cite{9756596, pmlr-v119-chan20a}.

\noindent \emph{Sparsity:} Many of the aforementioned works implement dense BNNs, where all parameters are used to create ensembles. To improve ensembling efficiency, theoretical studies have explored partial Bayesian inference, which involves sampling only select parameters or layers for ensembles \cite{sharma2023bayesian,kristiadi2020being, prabhudesai2023lowering}. In binary networks,  increased sampling sparsity (fewer probabilistic synapses) has been reported as training progresses \cite{Meng_Bachmann_Khan_2020}  due to greater predictor confidence and reduced uncertainty for in-distribution (IND) data \cite{Simeone_2022}. In hardware, \cite{bonnet2023bringing} performs ensembling only in the fully connected (FC) layers after feature extraction by the convolutional layers. In \cite{lu2022algorithm} only the deepest layers were sampled, which improved hardware efficiency but resulted in a larger number of ensembles for the deepest layers, and a drop in accuracy was observed.

\noindent \emph{Contributions:}
The main contributions of this work as:
\begin{itemize}
    \item We propose a sparsity-aware PCM-based in-memory computing (IMC) architecture optimization for Bayesian Binary Neural Networks (BBNNs) that leverages both layer sparsity (LS)—sampling only deep layers—and row sparsity (LS+RS)—sampling only rows with probabilistic synapses (Fig. \ref{fig:sparsity-intro}).
    Simulation results show no loss in accuracy or uncertainty quantification compared to the FP32 baseline, and a simple drift compensation technique maintains performance stability up to $10^7$\,s.
    \item We use the NeuroSIM \cite{peng2019dnn+} simulator to project hardware efficiency, and achieve up to $5.3\times$ smaller area, $8.8 \times$ power efficiency and $12.5 \times$ total efficiency with LS only and $45 \times$ with LS+RS, compared to non sparsity-aware implementation. We also achieve $2.9 \times$ higher power efficiency compared to the state-of-the-art.
\end{itemize}


\section{Background}\label{sec: Background}

\noindent \emph{Bayesian Neural Networks:} \label{subsec: training}
Given a BBNN parametrized by $\boldsymbol{w} \in \{-1,+1\}^{|\boldsymbol{w}|}$, Bayesian learning involves estimating the posterior $p(\boldsymbol{w}|\mathcal{D})$ which quantifies the likelihood of $\boldsymbol{w}$ under the evidence presented by the training dataset $\mathcal{D} = \{x_i, y_i\}_i$. The trained weights then follow a posterior distribution $p(\boldsymbol{w}|\mathcal{D})$ rather than a single point-estimate \cite{9756596}.

Bayesian learning in BBNNs is performed through mean-field variational inference (VI) described in \cite{Meng_Bachmann_Khan_2020} to obtain a collection of synaptic probabilities $\boldsymbol{p}_{\boldsymbol{w}} = \{p_w\}_{w \in \boldsymbol{w}}$ such that $p_w=Pr(w=+1)$. An ensemble of networks can be then created by sampling $\boldsymbol{w}$ (Fig.~\ref{fig:sparsity-intro}) which can be  accumulated  during inference  to obtain the final marginal predictor 
\begin{equation}
    p(y|\boldsymbol{x},\mathcal{D})
    = \frac{1}{N_{MC}}\sum_{i=1}^{N_{MC}} p(y|\boldsymbol{x},\boldsymbol{w}_i).
    \label{eq:ensembling}
\end{equation}
 Here $N_{MC}$ refers to the number of sampled predictors combined in Monte Carlo fashion.


\noindent \emph{Aleatoric and Epistemic Uncertainty:}
The total uncertainty of the prediction $U_{tot}$ for a given input $x$ is calculated as prediction entropy as specified in \cite{bonnet2023bringing}.  $U_{tot}$ is then decomposed into aleatoric uncertainty $U_a$ quantifying the uncertainty emerging from the measurement noise of in-distribution (IND) data, and epistemic uncertainty $U_e$ quantifying the inherent uncertainty in the model caused by limited evidence due to inadequate amount of data $\mathcal{D}$.  

When all individual predictors $p(y | \boldsymbol{x},\boldsymbol{w})$ agree, $U_e=0$. Upon disagreement, it is positive indicating an out-of-distribution (OOD) input. Note that for traditional non-Bayesian frequentist networks with point estimate weights $U_{tot}=U_a$, and therefore $U_e=0$.

\noindent \emph{PCM Devices:} A phase change memory (PCM) device consists of a chalcogenide material such as Ge$_2$Sb$_2$Te$_5$ sandwiched between two metal electrodes.  The device conductance is programmed using a write-and-verify scheme varying the programming pulse magnitude iteratively \cite{doi:10.1063/1.5042408,rajendran2019building}. PCM devices exhibit Gaussian-distributed programming and read noise with the former having a dominant effect, as well as a state-dependent drift in programmed conductance levels. 

\section{Sparsity-aware PCM IMC Architecture}\label{sec:architecture}
In this section, we introduce the proposed sparsity-aware IMC optimization for PCM-based hardware architecture. We discuss the sampling sparsity in network and introduce LS and LS+RS schemes for BBNN accelerator architecture.


\begin{figure}
\centering

    \includegraphics[width=0.5\textwidth]{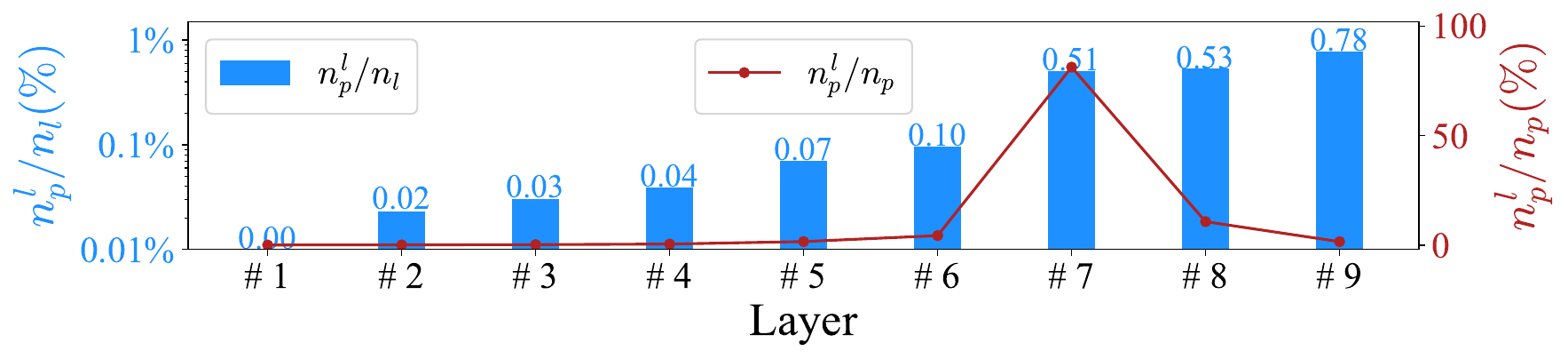}%
\caption{Probabilistic synaptic parameter concentration across layers. The bar plot depicts the fraction of the layer that is probabilistic ($n^l_p/n^l$). The line plot shows the layerwise proportion of all probabilistic synapses ($n^l_p/n_p$).} 
\label{fig:sparsity}
\end{figure}

\noindent\emph{Sparsity}:  For a binary-valued synapse $w \in \boldsymbol{w}$ with $p_w = \mathrm{Pr}(w=1) $, $p_w=0$ indicates that the synapse is always $-1$, while $p_w=1$ implies that it is $+1$. The extreme cases are deterministic synapses with fixed values across all samples, whereas the other synapses, that are probabilistic, take either value depending on $p_w$. Let $n_d$ the total number of deterministic synapses and $n_p$ total number of probabilistic synapses be $n_p$, with $n_p^l$ denoting the number of probabilistic synapses in layer $l$ and $n^l$ the total number of synapses in that layer. Fig.~\ref{fig:sparsity}  shows the fraction of stochastic weights in each layer ($n^l_p/n^l$), and the share of total probabilistic synapses per layer ($n^l_p/n_p$) for VGGBinaryConnect trained on CIFAR-100. We observe significant sampling sparsity in the network, i.e., $n_d \gg n_p $ \cite{Meng_Bachmann_Khan_2020}. To further increase sparsity, $p_w$ near the extreme values were clamped to $0$ or $1$.

\begin{figure}
\centering
    \includegraphics[width=0.45\textwidth]{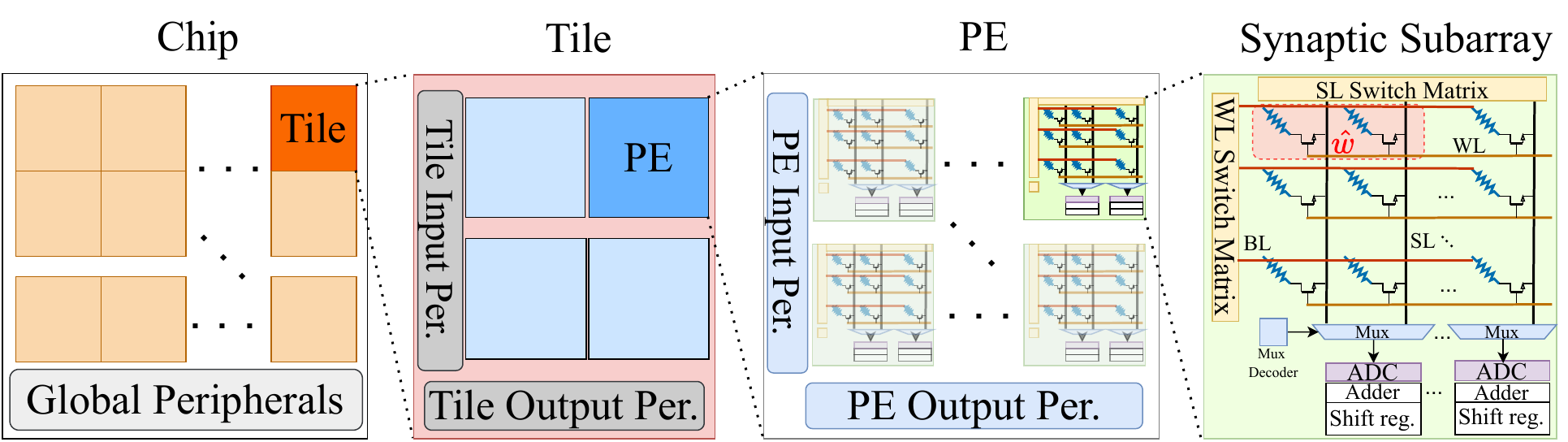}
\caption{The accelerator chip architecture, based on \cite{peng2019dnn+}. } 
\label{fig:B-SpaNN}
\end{figure}
   
\noindent\emph{Chip Heirarchy}: Fig.~\ref{fig:B-SpaNN} shows the hierarchical structure of the NeuroSIM accelerator   architecture  \cite{peng2019dnn+}. The chip contains compute tiles, along with global buffers, accumulation, batch normalization (BN), and pooling units. Each tile maps weights from only a single layer, but multiple tiles can be used if the layer exceeds the tile size. Tiles are divided into processing elements (PEs), which house 1T-1R PCM crossbar synaptic subarrays. 

\begin{figure}
\centering
\begin{subfigure}{0.15\textwidth}
\centering
    \includegraphics[width=\textwidth]{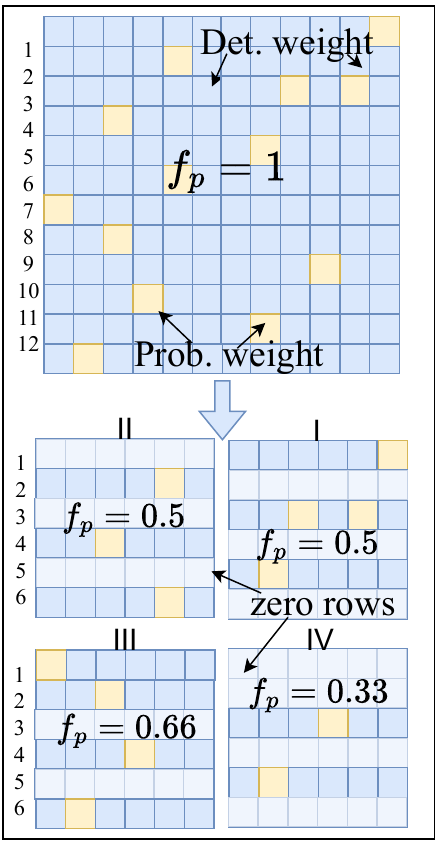}
\end{subfigure}
\begin{subfigure}{0.25\textwidth}
\centering
    \includegraphics[width=\textwidth]{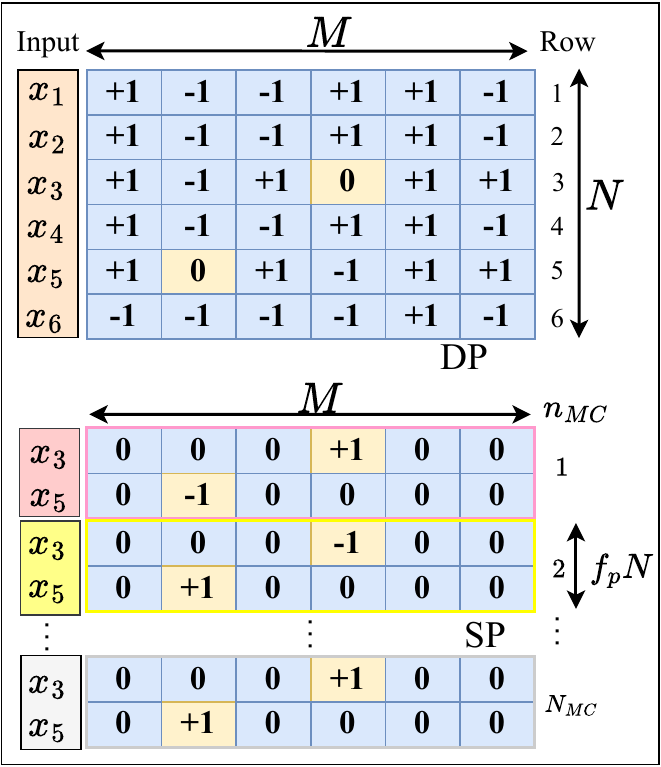}
\end{subfigure}
\caption{\textbf{Left}: Partitioning matrix into $\boldsymbol{W}^D$ and $\boldsymbol{W}_i^S$. Yellow-colored cells represent probabilistic synapses, while blue ones represent deterministic. \textbf{Right}: DP-SP split and MVM operation for submatrix IV. For an $f_p=0.33$, the $N_{MC}=10$ submatrices will require $4$ subarrays.  Inputs corresponding to non-zero rows are sliced and fed to each $\boldsymbol{W}_{i,sub}^S$ of the SP.   } 
\label{fig:sparsity_splitting}
\end{figure}

\noindent\emph{Layer Sparsity (LS)-only scheme}: Each probabilistic layer generates $N_{MC}$ outputs according to \eqref{eq:ensembling}, requiring multiple instantiations in subsequent layers regardless of sparsity. To minimize complexity, we select an appropriate first ensembling layer (FEL) with significant probabilistic synapses, ideally deep in the network. Based on Fig.~\ref{fig:sparsity}, layer 7 is chosen as the FEL for this network. Synapses in earlier layers are clamped to 0 or 1, while only layers 7-9 are ensembled in space, with all $N_{MC}$ samples processed in parallel.

 \noindent\emph{DP-SP Architecture}: To further increase area efficiency we exploit the inherent probabilistic sparsity in these layers by dividing and mapping them into a deterministic plane (DP) and the stochastic plane (SP). All the rows containing probabilistic synapses are extracted into sub-matrices and sampled to create ensembles. 
 To generate the layer output for ensemble $i$, the DP outputs are combined with the output $i^{th}$ submatrix in SP, as illustrated in Fig.~\ref{fig:sparsity-intro}. 
 
To illustrate the splitting mechanism shown in Fig.~\ref{fig:sparsity_splitting}, consider mapping a layer of the $i^{th}$ network sample with weight matrix  $\boldsymbol{W_i}$, with the corresponding DP and SP counterparts represented as $\boldsymbol{W}_i^D$ and $\boldsymbol{W}_i^S$. Each element $\hat{w} \in \boldsymbol{W}_i^D$ is
\begin{equation}
    \hat{w} = 
     \begin{cases}
        w &\quad \text{if } p_w \in \{0,1\},\\
       0 &\quad otherwise.\\
     \end{cases}
    \label{eq:dp}
\end{equation}
Thus $\boldsymbol{W}_i^D=\boldsymbol{W}_j^D=\boldsymbol{W}^D$ for all $i \neq j$. The probabilistic synapses are contained in $\boldsymbol{W}_i^S$ such that $\boldsymbol{W_i} = \boldsymbol{W}^D + \boldsymbol{W}_i^S$ and an MVM operation for an input $\boldsymbol{x}$ is given by 
 \begin{equation}
     \boldsymbol{y_i} = \boldsymbol{W}_i\boldsymbol{x} = \boldsymbol{W}^D\boldsymbol{x} +  \boldsymbol{W}_i^S\boldsymbol{x}.
     \label{sparse_split}
 \end{equation}

  As $\boldsymbol{W}_i^S$ is a sparse matrix, rows with no non-zero values can be removed before mapping. However, we observe that only a few rows of $\boldsymbol{W}_i^S$ are zero rows, as illustrated in Fig.~\ref{fig:sparsity_splitting}, even though the rows themselves are sparse vectors. Empirically, we observed that while the minimum sparsity in any row vector was around $98\%$, only $1.5\%$ of all rows were zero rows. However, as each $\boldsymbol{W}_i^S$ needs to be split into submatrices corresponding to subarray dimensions before mapping, the smaller submatrices $\boldsymbol {W}^S_{i,sub}$, have sufficient row sparsity (RS).  The sparsity increases with decreasing subarray size. Hence, we can eliminate the zero rows and squeeze each submatrix before mapping to hardware. For example, if the fraction of non-zero rows in a submatrix,  $f_p=0.1$, then $10$ ensembles can fit into one subarray. The fraction $f_p$ is calculated for each subarray mapping to optimize packing. 
 
To perform the MVM operation, an input subvector from $\boldsymbol{x}$, corresponding to the non-zero row indices, is fed to the SP subarrays. Column sums accumulated in each local buffer (PE/tile) are then globally combined to obtain the final pre-neurons, which are passed to global BN, ReLU, and Pooling units. While column sparsity exists at the subarray level, squeezing columns causes misalignment, preventing parallel row readouts, and requiring extra circuitry and a more complex sparsity handling strategy \cite{yue2024scalable}.
 
For maximum area efficiency, both LS and RS schemes can be combined into LS+RS. With multiple ensembles packed into the same subarrays, PEs, and tiles, ensembling is done sequentially, unlike the parallelism in LS-only scheme.

\begin{figure}
  \centering
    \includegraphics[width=0.4\textwidth]{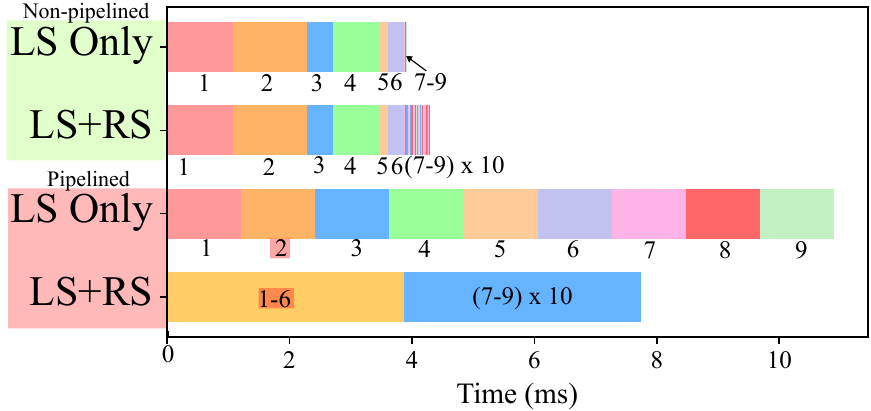}
    \caption{Timing analysis for pipelined and non-pipelined mode for both LS and LS+RS modes. $\times 10$ highlights the $10$ ensembles that are processed sequentially. Stage numbers highlighted in red in pipelined mode decide stage latency.}
  \label{fig:timing}
\end{figure}

\noindent\emph{Timing}: Fig.~\ref{fig:timing} illustrates the timing strategy for both LS and LS+RS strategies in pipelining and non-pipelining modes. As discussed earlier, in the LS+RS scheme, predictions are available sequentially, trading latency for area efficiency compared to the LS-only scheme. However, we note that shallower convolutional layers require longer processing times than an FC layer due to the `slide-over' operation of convolutional kernels and larger input feature maps \cite{peng2019dnn+} making additional latency per image minimal.

In pipelined LS-only mode, each layer functions as a pipeline stage, with stage latency set by the slowest layer, minimizing control overhead. In LS+RS mode, where layers 7-9 are processed serially, the same pipelining strategy would create stalls, as shallower layers would need to pause for ensemble processing in deeper layers. To address this, we propose a two-stage pipeline: layers 1-6 process in stage 1, and all ensembles for layers 7-9 complete sequentially in stage 2. For $N_{MC} = 10$, the shallow layers dominate the latency, so the two-stage LS+RS pipeline minimizes per-image latency, though the LS-only mode provides better throughput in pipelined operation due to reduced stage latency.

\noindent\emph{Weight mapping, operation, and drift compensation}: Each weight of the network $w\in \{-1,1\}$.  In addition, many cells will be mapped to $0$ as per \eqref{eq:dp}. To avoid non-linearity due to drive voltage\cite{chen2015technology}, row inputs are sequentially bit-streamed and the products are shift-added at every column for $n$ cycles, where $n$ is the input bit-width. Therefore, each synapse requires 2-bit storage resulting in two devices per cell, as shown in Fig.~\ref{fig:B-SpaNN}. The weight-to-conductance scaling is chosen to utilize the full dynamic range of conductances available for PCM devices (0-25 $\mu$S).

As discussed earlier, the devices experience conductance drift with a state-dependent exponent $\nu$. Since all devices store conductance values of either $0$ or $25\mu$S, the weights drift with the same nominal value of the exponent, reducing all pre-neuron values by approximately the same factor. To counter this, we periodically scale the BN coefficients to restore the pre-neurons to their nominal values. Unlike the common practice of subsuming BN coefficients into the synaptic layer by scaling the synapses \cite{perez2021heterogeneous}, we apply BN layer separately to preserve the ternary nature of the weights.  
\vspace{-.1in}
\section{Experiments and conclusions} \label{sec:results}
\begin{figure}
\centering    
        \includegraphics[width=0.5\textwidth]{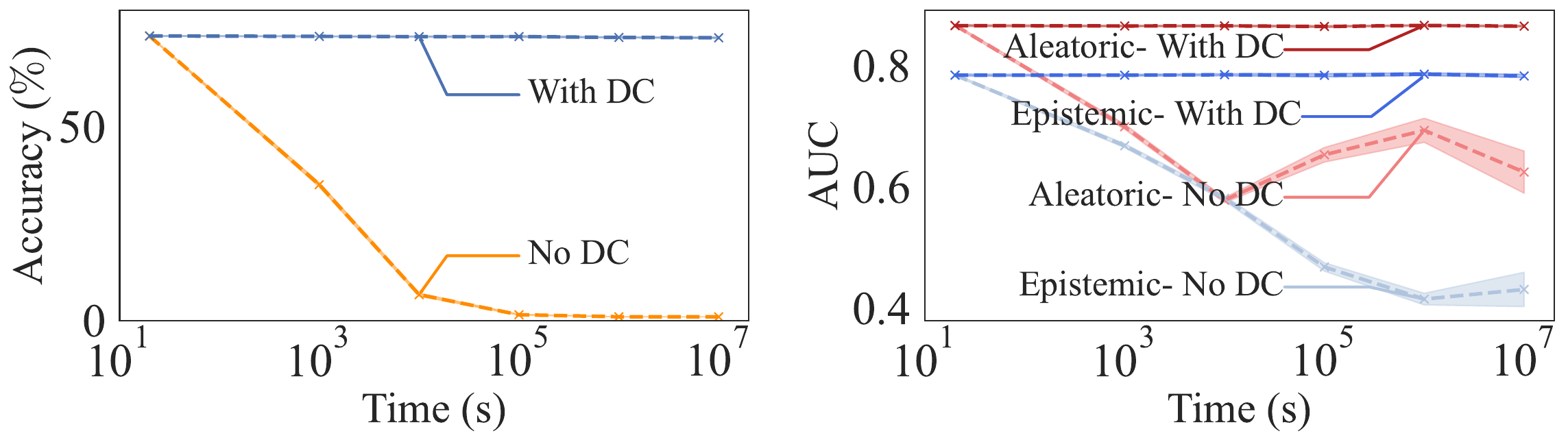}
        \caption{Accuracy (Left) and AUC of ROC curves for aleatoric and epistemic uncertainty (Right) for CIFAR-100 dataset. The band around lines represent one standard deviation. }
        \label{accuracy}
\end{figure}

We present the accuracy and uncertainty results on CIFAR-100 dataset with VGGBinaryConnect network transferred to a custom PCM simulator that incorporates device characteristics from \cite{joshi2020accurate} and the sparsity characteristics outlined here. We assess classification accuracy, uncertainty quantification, and stability under drift (Fig.~\ref{accuracy}). Additionally, we estimate area, energy, and latency using the NeuroSIM emulator \cite{peng2019dnn+}.

\noindent\emph{Accuracy and Uncertainty Performance}: To evaluate epistemic uncertainty, we use CIFAR-10 as the OOD data, given its mutually exclusive classes from CIFAR-100 \cite{krizhevsky2009learning}. Aleatoric and epistemic uncertainties are assessed using the AUC of the ROC curve. Aleatoric ROC specifies the ability to differentiate correct from incorrect IND data, while epistemic ROC characterizes IND vs. OOD discrimination capability \cite{bonnet2023bringing}. 

We see minimal accuracy drop compared to the FP32 baseline (73.34\% vs. 73.68\%) despite clamping a few shallow-layer parameters. Drift compensation (DC) by scaling BN parameters completely recovers accuracy and aleatoric uncertainty performances for up to $10^7$ seconds.

\noindent\emph{Hardware Projection}: We estimate hardware performance by evaluating PCM IMC accelerator described here with  $128 \times 128$ subarrays on 90 nm technology, a node chosen to match device characteristics in \cite{joshi2020accurate}. Throughput in pipelined mode is measured in frames per second (FPS), with power and total efficiency in FPS/W and FPS/W/cm\(^2\) respectively. In non-pipelined mode, latency measures compute speed, with the energy-delay product (EDP) and latency-energy-area product (LEAP) used as figures of merit (FoMs). We achieve $2.9 \times$ higher power efficiency than the state-of-the-art reported in \cite{lu2022algorithm}. Table \ref{tab:comparison} shows hardware estimates. Both LS and LS+RS modes are around $8.8\times$ more power-efficient than non-sparsity-aware implementations. LS+RS mode, with $3.7\times$ smaller area, achieves $3.8\times$ higher total efficiency than LS mode alone. In non-pipelined mode, LEAP improves by $12.5\times$ in LS mode, with a further $3.6\times$ gain in LS+RS.

\begin{figure}
\centering
        \includegraphics[width=0.5\textwidth]{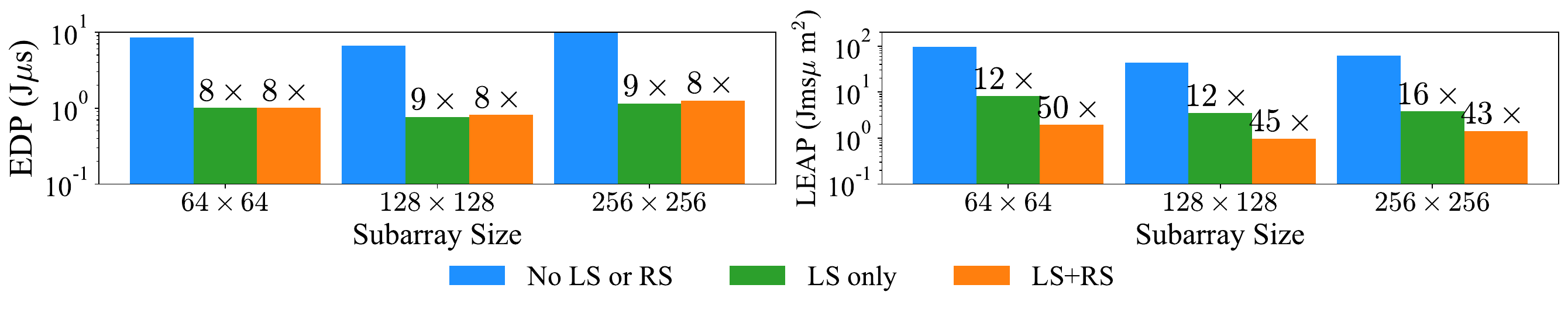}    
    \caption{EDP and LEAP for non-sparsity aware, LS only and LS+RS sparsity schemes. The labels on top of the bars indicate improvement over the non-sparsity-aware implementation.}
    \label{fig:LEA}
\end{figure}

Subarray size also impacts performance. Smaller subarrays reduce latency but at the cost of area and energy efficiency due to a smaller IMC-to-peripheral area ratio. Fig.~\ref{fig:LEA} shows significant FoM improvements with sparsity. While EDP gains are minimal from LS to LS+RS, area reduction in RS results in a $3-4\times$ increase in LEAP. We find $128 \times 128$ to be optimal vis-\`a-vis both FoMs.

\begin{table}
    \centering
\caption{Hardware performance of various sparsity schemes}
\label{tab:comparison}
    \begin{tabular}{cccc} 
    \hline
         \textbf{Parameter}&  \textbf{No LS or RS}&  \textbf{LS only}& \textbf{LS+RS}\\
         \hline\hline
         \textbf{Pip.latency (per stage; ms)}&  1.43&  1.43& 4.00\\ 
         \textbf{Energy (mJ)/image}&  1.51&  0.17& 0.17\\ 
         \textbf{Area (cm$^2$)}&  65.38&  45.09& 12.2\\ 
         \textbf{Throughput (FPS)}&  698&  698& 250\\ 
         \textbf{Power Eff. (FPS/W)}&  664&  5722& 5938\\ 
         \textbf{Total Eff. (Pip; FPS/W/cm$^2$)}&  10.2&  126.9& 488\\ 
         \textbf{Non-Pip. latency (ms)}&  4.02&  4.02& 4.29\\
         \textbf{EDP (J$\mu$\,s)}&  6.66&  0.77& 0.8\\
         \textbf{LEAP (Jms$\mu$m$^2$)}&  43.53&  3.48& 0.97\\
         \hline
    \end{tabular}
\end{table}
\vspace{-.1in}
\section{Conclusion}
In this work, we presented two sparsity-aware BNN acceleration schemes: LS, which addresses layer-wise sampling sparsity, and RS, which exploits row sparsity. We achieved competitive accuracy on a PCM simulator, demonstrated uncertainty quantification, and effectively detected OOD samples. We also validated a simple drift compensation method completely recovering drift degradation of performance for up to $10^7$ seconds. Our approach improves power efficiency by $8.8\times$ and total efficiency by $12.5\times$ (LS) and $45\times$ (LS+RS). Similar gains were observed in non-pipeline mode. 

We explored hardware efficiency trade-offs based on subarray size, finding $128 \times 128$ to be optimal. While results may vary with different architectures and input sizes, the proposed schemes are adaptable to any convolutional or other BNN implementation. \newpage
\ignore{
\label{sec:conc} 
\section*{Acknowledgment}
  This research was supported by the EPSRC Open Fellowship EP/X011356/1 and the EPSRC grant EP/X011852/1.}

\bibliography{biblio_aicas}
\end{document}